\documentclass[11pt,tightenlines,showpacs]{revtex4}
\usepackage{graphicx}
\usepackage{float}
\usepackage[T1]{fontenc}
\usepackage[latin1]{inputenc}
\usepackage{amssymb}
\newcommand{\be}{\begin{equation}}
\newcommand{\ee}{\end{equation}}
\newcommand{\bea}{\begin{eqnarray}}
\newcommand{\eea}{\end{eqnarray}}

\makeatother
\begin{document}

\preprint{TUW-09-07}

\title{On the imaginary part of the next-to-leading-order static gluon self-energy\\
in an anisotropic %non-Abelian 
plasma
}

\author{M.E. Carrington}

\affiliation{Brandon University,
Brandon, Manitoba, R7A 6A9 Canada\\
and Winnipeg Institute for Theoretical Physics,
Winnipeg, Manitoba, Canada}

\author{A. Rebhan}

\affiliation{Institut f\"{u}r Theoretische Physik, Technische Universit\"{a}t Wien, Wiedner Hauptstrasse 8-10, A-1040 Vienna, Austria}

\date{September 14, 2009}

\begin{abstract}
Using hard-loop (HL) effective theory for an anisotropic non-Abelian plasma, which even in the static limit involves nonvanishing HL vertices,
we calculate the imaginary part of the static next-to-leading-order
gluon self energy
in the limit of a small anisotropy and with
external momentum parallel to the anisotropy direction.
At leading order, the static propagator has space-like poles corresponding
to plasma instabilities.
On the basis of a calculation
using bare vertices, it has been conjectured that,
at next-to-leading order, the static gluon self energy
acquires an imaginary part which regulates these space-like poles.
We find that the one-loop resummed expression taken over
naively from the imaginary-time formalism does yield
a nonvanishing imaginary part 
even after including all HL vertices. 
However, this result is not correct. Starting from
the real-time formalism, which is required in a non-equilibrium
situation, we construct a resummed retarded HL propagator
with correct causality properties and
show that the static limit of the retarded
one-loop-resummed gluon self-energy is real.
This result is also required for the time-ordered propagator
to exist at next-to-leading order.
\end{abstract}


\maketitle

\section{Introduction}

Non-Abelian plasma instabilities \cite{Mrowczynski:1988dz,Pokrovsky:1988bm,Mrowczynski:1993qm} have received a lot of attention following the apparent
failure of perturbative QCD at finite temperature to account for the
rapid thermalization and strong collectivity deduced from the experimental
results at the Relativistic Heavy Ion collider (RHIC) \cite{Tannenbaum:2006ch}. To leading order (LO) in the coupling and for small gauge field amplitudes, the dynamics of plasma instabilities are determined by the generalization of the hard-thermal-loop (HTL) \cite{Weldon:1982aq,Frenkel:1990br,Braaten:1990mz} gauge boson self-energy to anisotropic situations \cite{Mrowczynski:2000ed,Romatschke:2003ms,Romatschke:2004jh,Arnold:2003rq}.
At LO, the scale associated with plasma instabilities is of the same
parametric order as that of other collective phenomena 
such as the Debye mass, and consequently plasma instabilities strongly modify previous
perturbative thermalization scenarios
\cite{Arnold:2003rq,Bodeker:2005nv,Mueller:2006up}.

Much progress has been made in
the study of the nonlinear evolution of
non-Abelian plasma instabilities using real-time lattice simulations \cite{Rebhan:2004ur,Arnold:2005vb,Rebhan:2005re,Bodeker:2007fw,Arnold:2007cg,Rebhan:2008uj} using
hard-loop (HL) effective theory  \cite{Mrowczynski:2004kv}
(for classical field theory simulations see \cite{Romatschke:2005pm,Romatschke:2006nk,Berges:2007re,Berges:2008zt}).

Also a number of analytical calculations have been performed.
The leading order anisotropic gluon propagator has been analysed in the temporal axial gauge in \cite{Romatschke:2003ms,Romatschke:2004jh,Arnold:2003rq}, and in covariant gauge in \cite{Dumitru:2007hy,Guo:2008da} and \cite{Carrington:2008sp}. Unlike the HTL gluon propagator, this anisotropic propagator contains non-integrable space-like poles, which signal the presence of instabilities in an anisotropic system. 
The HL gluon propagator has been used for perturbative studies of
collisional energy loss \cite{Romatschke:2003vc,Romatschke:2004au} and the decay width of quarkonium bound states \cite{Burnier:2009yu,Dumitru:2009fy} in an
anisotropic plasma, where these space-like poles do not present
a fundamental problem. However,
non-integrable singularities at zero frequency occur in perturbative calculations of jet quenching and momentum broadening in the anisotropic quark-gluon plasma \cite{Romatschke:2006bb,Baier:2008js}, where they appear to signal an enhancement of these observables compared to the equilibrium case.
It was suggested in \cite{Romatschke:2006bb} that these singularities are regulated by a non-zero imaginary part of the next-to-leading order (NLO)  gluon polarization tensor. A partial one-loop result using the HL-resummed propagator but bare vertices was performed, suggesting that this was indeed the case.

In this paper we shall show that the one-loop resummed expressions considered in \cite{Romatschke:2006bb} do indeed yield a nonvanishing imaginary part after adding up all HL-resummed one-loop contributions including the nontrivial vertices present even at zero frequency. However, starting from the real-time formalism we show that these one-loop resummed expressions need to be evaluated using the static limit of the retarded propagator, but in the presence of plasma instabilities the retarded propagator is no longer obtained from the analytic HL propagator by a limiting procedure. Constructing and employing a resummed retarded HL propagator with correct causality properties we find that the NLO static gluon self-energy is real.

In thermal equilibrium, the imaginary part of the static gluon self energy would be required to vanish due to the KMS conditions \cite{LeB:TFT}. In \cite{Romatschke:2006bb} it was argued that, if the KMS conditions are violated in the anisotropic case, there might be a finite, discontinuous contribution.
We note, however, that such a nonvanishing imaginary part to the self energy in the static limit would cause
a fundamental problem with real-time perturbation theory, because 
the static limit of the time-ordered HL-propagator would become ill-defined. This can be seen as follows. In the limit of small anisotropies, we can use equilibrium distribution functions on soft HL-resummed lines, and therefore the time-ordered propagator is obtained from the retarded propagator using
\cite{Beraudo:2007ky}
\bea
D^{T}(k_0,k) = D^{ret}(k_0,k)+2 \,n(k_0)\, i \,{\rm Im}\,D^{ret}(k_0,k),
\eea
where $n(k_0)$ is the Bose distribution function.
The second term in this expression diverges for $k_0\rightarrow 0$, if the imaginary part of the static self-energy is non-zero.

\begin{figure}[H]
\begin{center}
\includegraphics[width=10cm]{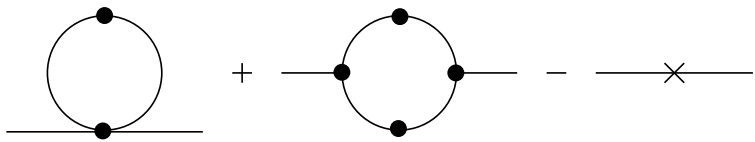}
\end{center}
\caption{The diagrams that contribute to static gluon self energy. All lines correspond to gluon propagators. The dots indicate hard loop propagators and vertices. The cross denotes the counterterm.}
\label{staticSE}
\end{figure}

The anisotropic NLO static gluon self-energy is obtained from the diagrams shown in Fig. \ref{staticSE}. In equilibrium,
only static loop momenta have to be considered and no fermion loops contribute \cite{Rebhan:1993az} at next-to-leading order, and we shall see that this is also true in the anisotropic case.
Since the ghost self energy vanishes at leading order, the ghosts do not need to be resummed. The solid dots indicate leading order propagators and vertices which are obtained from the HL effective action \cite{Mrowczynski:2004kv}. In contrast to the equilibrium situation, the HL vertices do not reduce to bare vertices in the static limit. The third diagram represents the HL counterterm which must be subtracted to avoid double counting. 
%%%%%%%%%%%%%%%%%%%
\par
%%%%%%%%%%%%%%%%%%%%

At zero temperature, field theory can be formulated covariantly. At finite temperature, covariance is broken by the vector $u_\mu=(1,0,0,0)$ which specifies the rest frame of the thermal system. For anisotropic systems we need (in the simplest case) one additional vector to specify the direction of the anisotropy. We consider the case in which there is one preferred spatial direction along which the system is anisotropic (in planes transverse to this vector the system is isotropic). In the context of heavy ion collisions, we can take this direction to be the beam axis ($\hat z$) along which the initial expansion occurs.

In \cite{Carrington:2008sp} we have given the complete analytic result for the integrand corresponding to the diagrams in Fig. \ref{staticSE}. Although this result is relatively compact, the evaluation of the integrals would be a formidable task.
In order to simplify the calculation, we choose a particular orientation for the external momentum. We calculate the part of the NLO static gluon self energy related to Weibel instabilities for an anisotropic plasma, in the limit of small anisotropy, with the external momentum parallel to the direction of the anisotropy. 

This paper is organized as follows. In section \ref{sectionNot} we define our notation. In section \ref{sectionIntegrand} we present the calculation of the integrand. In section \ref{sectionProp} we discuss the analytic structure of the HL propagator. In section \ref{sectionNaive} we calculate the NLO contribution using the analytic propagator obtained by resumming the HL gluon self-energy. 
In section \ref{sectionMod} we construct a retarded HL-resummed propagator and identify an extra contribution that cancels the imaginary part obtained in section \ref{sectionNaive}. In section \ref{sectionConclusions} we present our conclusions.

\section{Notation and ingredients}
\label{sectionNot}

Throughout this paper we will frequently use the indices $\{k,q,r\}$ to denote momentum arguments. We also use Latin letters $\{i,j,l,\cdots\}$ to denote spatial indices, with the exception that the indices $k$, $q$, $r$ are reserved and used exclusively to denote spatial momenta. Four-momenta will be denoted by
capital letters.
The external momentum in Fig. \ref{staticSE} will be called $\vec q$, and the internal momenta are $\vec k$ and $\vec r=-\vec k-\vec q$. We choose $\vec q = (0,0,q)$.

\subsection{Distribution Functions}

%We define the isotropic distribution:
For a non-Abelian plasma with $N_c$ gluons and $N_f$ quark flavors,
we define an effective isotropic distribution function by
\bea
\label{f-iso}
f_{\rm iso}(p)=2 N_f[n(p)+\tilde n(p)]+4 N_c n^g(p).
\eea
In thermal equilibrium, we have
\bea
n_{\rm eq}(p)=\frac{1}{e^{(p-\mu)\beta}+1}\,,~~\tilde n_{\rm eq}(p)=\frac{1}{e^{(p+\mu)\beta}+1}\,,~~n^g_{\rm eq}(p)=\frac{1}{e^{p\,\beta}-1}.
\eea
We define the Debye mass from the equilibrium distribution$^1$
\footnotetext[1]{In Ref. \cite{Romatschke:2004jh} Eq. (\ref{debye}) differs by a factor of 2 and Eq. (\ref{f-iso}) differs by a factor 1/2. The definition of the Debye mass is the same.}
\bea
\label{debye}
m_D^2=g^2 \int \frac{d^3 p}{(2\pi)^3} \,\frac{f_{eq}(p)}{p} = -\frac{g^2}{2} \int \frac{d^3 p}{(2\pi)^3} \, \frac{d f(p)}{d p} = \frac{1}{3}N_c g^2 T^2+\frac{1}{6}N_f g^2 \left(T^2+\frac{3}{\pi^2}\;\mu^2\right)\,.
\eea

Following \cite{Romatschke:2004jh}, we can construct an anisotropic distribution from any isotropic distribution of the form $f_{\rm iso}(p^2)$ by writing
\bea\label{fxi}
f(\vec p) = f_{\rm iso}(\sqrt{p^2+\xi(\vec p\cdot \hat z)^2}),
\eea
where $\xi > -1$ is the anisotropy parameter. A value $\xi>0$ corresponds to a contraction of the distribution and $0>\xi>-1$ corresponds to a stretching of the distribution. In the following we shall consider only deformations of the equilibrium distribution. For nonzero $\xi$, the parameters $T$ and $\mu$ of course lose the usual meaning of temperature and chemical potential. In this paper we restrict ourselves to weakly anisotropic systems for which $|\xi|\ll 1$ and shall calculate only to leading order in $\xi$.

\subsection{Vertices}

Momenta are taken to be incoming. We give only the tadpole form of the 4-point vertex, since that is the only 4-point vertex we will need. The bare vertices are:
\bea
&& (\Gamma_0)_{abc}^{\mu\nu\lambda} = igf_{abc}\Gamma_0^{\mu\nu\lambda} \\
&& \Gamma_0^{\mu\nu\lambda} = -g^{\mu\nu}(K^\lambda-Q^\lambda)-g^{\lambda\nu}(Q^\mu-R^\mu)-g^{\lambda\mu}(R^\nu-K^\nu) \nonumber\\
&& (M_0)_{abcc}^{\mu\nu\lambda\sigma}(Q,-Q,K,-K) = 2 g^2\,C_A \delta_{ab} \;M_0^{\mu\nu\lambda\sigma} \nonumber\\
&& M_0^{\mu\nu\lambda\sigma} = -g^{\lambda\nu}g^{\mu\sigma}-g^{\lambda\mu}g^{\nu\sigma}+2g^{\lambda\sigma}g^{\mu\nu}\nonumber
\eea

The notation for the HL vertices is given in \cite{Mrowczynski:2004kv,Carrington:2008sp}. We define:
\bea
&&\int_p:= \frac{d^3p}{(2\pi)^3} \Big|_{p_0=p}\;;~~ \hat P^\mu := (1,\hat p^i)\;;~~
\hat I_\beta := \frac{g^2}{2}\int_p \frac{\partial f}{\partial P^\beta}
\eea
The 2-point function is:
\bea
&& \Pi_{ab}^{\mu\nu}:=\delta_{ab}\Pi^{\mu\nu} \;;~~\Pi^{\mu\nu} := \hat I_\beta \hat P^\mu \, \left(g^{\nu\beta}-\frac{\hat P^\nu Q^\beta}{P\cdot Q}\right)
\eea
The 3-point function is:
\bea
\label{GammaDef}
&&\Gamma^{\mu\nu\lambda}_{abc}:=i g f_{abc}\Gamma^{\mu\nu\lambda} \\
&& \Gamma^{\mu\nu\lambda} := \hat I_\beta P^{\mu}P^{\nu}P^{\lambda}\;
\left(\frac{K^\beta}{\hat P\cdot K\;\hat P\cdot Q} - \frac{R^\beta}{\hat P\cdot Q\;\hat P\cdot R}\right)\nonumber
\eea
We need only the tadpole form of the 4-point vertex which has the form:
\bea
\label{MDef}
&& M^{\mu\nu\lambda\sigma}_{abcc}(Q,-Q,K,-K):=2 g^2\,C_A \delta_{ab} \;M^{\mu\nu\lambda\sigma}(Q,-Q,K,-K) \\
&&M^{\mu\nu\lambda\sigma}(Q,-Q,K,-K) := -2 \hat I_\beta \hat P^\mu \hat P^\nu \hat P^\lambda \hat P^\sigma \left(\frac{K^\beta \hat P \cdot Q- Q^\beta \hat P \cdot K }{P \cdot K\;P \cdot Q\;((P \cdot K)^2-(P \cdot Q)^2)}\right) \nonumber
\eea
These vertices satisfy the Ward identities:
\bea
\label{wi-full}
&& K_\mu \Gamma^{\mu\nu\lambda}(K,Q,R) = \Pi^{\nu\lambda}(Q) - \Pi^{\nu\lambda}(R) \\
&&K_\lambda M^{\mu\nu\lambda\sigma}(Q,-Q,K,-K) = -2 \Gamma^{\mu\nu\sigma}(K,Q,-K-Q) \nonumber\\
&&K_\lambda K_\sigma M^{\mu\nu\lambda\sigma}(Q,-Q,K,-K) = 2 \Pi^{\mu\nu}(-K-Q) - 2 \Pi^{\mu\nu}(Q) \nonumber
\eea

\subsection{Leading order hard-loop self energies and propagators}
\label{sectionSE}

In Feynman gauge, which we use, the relation between
the polarization tensor $\Pi$ and full (bare) propagators $D$ ($D^0$)
is given by the relation
\bea
\label{PIdef}
D^{-1}_{\mu\nu}(K) =(D^{0}_{\mu\nu})^{-1}-\Pi_{\mu\nu} = -(g_{\mu\nu}K^2+\Pi_{\mu\nu}).
\eea
The HL gluon self-energy is gauge invariant and satisfies the usual Ward identity: $K^\mu \Pi_{\mu\nu}=0$. Consequently, we only need to calculate the spatial components.

In the isotropic (HTL) case, the spatial self energy has two independent components which are called the transverse and longitudinal parts.  Eq.~(\ref{PIdef}) defines an analytic propagator
from which the retarded (advanced) propagator is obtained as
\be\label{Dretadv}
D^{ret/adv}_{\mu\nu}(K)=D_{\mu\nu}(k_0\pm i\epsilon,\vec k).
\ee

For anisotropic systems the self energy can be generally decomposed into four independent structure functions. For the special case of $\Pi_{\mu\nu}(q_0,\vec q)$ with $\vec q = (0,0,q)$, there are only two independent structure functions.
We define the general projection operators using the vector \cite{Kobes:1991dc}
\bea
\label{n-defn}
n_k^i:=n^i(k)=(\delta^{ij}-k^ik^j/k^2)\delta^{j3}
\eea
which satisfies $n_k^i \,k^i = 0$.
Using this vector we construct the projection operators for $\Pi_{ij}(K)$:
\bea
\label{projAll}
&&P^{1k}_{ij}= \delta _{ij}-\frac{k_i k_j}{k^2}\,,~~
P^{2k}_{ij}=\frac{k_i k_j}{k^2}\,,~~
P^{3k}_{ij}= -\frac{k_i k_j}{k^2}-\frac{n^k_i n^k_j}{n_k^2}+\delta _{ij}\,,\nonumber\\
&&
P^{4k}_{ij}=k_j n^k_i+k_i n^k_j\,,~~
P^{5k}_{ij}=P^{1k}_{ij}-P^{3k}_{ij},
\eea
and similarly for $\Pi_{ij}(R)$.
The projection operators for $\Pi_{ij}(Q)$ are defined as:
\bea
P^{1q}_{ij}=\delta_{ij}-\delta_{3i}\delta_{3j}\,;~~P^{2q}_{ij}=\delta_{3i}\delta_{3j}
\eea
The orthogonality relations satisfied by these projectors are given in Appendix A. Using these definitions, the self energies can be decomposed as:
\bea
\label{P1q}
&&\Pi^q_{ij} = P^{1q}_{ij}\alpha_q + P^{2q}_{ij}\,\bar\beta_q \\
&&\Pi^k_{ij} = P^{1k}_{ij}\alpha_k + P^{2k}_{ij}\,\bar\beta_k + P^{5k}_{ij}\,\gamma_k + P^{4k}_{ij}\,\bar\delta_k\nonumber
\eea
The complete LO results for the HL self energy components $\alpha_k$, $\bar\beta_k$, $\gamma_k$, $\bar\delta_k$ have been given in \cite{Romatschke:2004jh} for the anisotropic distribution function (\ref{fxi}).

\subsection{Static propagator}

The static propagator has
(space-like) poles at $q_0=0$ and values of $\vec q$ which mark the boundaries
between stable and unstable modes.
These are determined by
the static parts of $\alpha_k$ and $\gamma_k$, which are given by:
\bea
\label{alphagammaRes}
\alpha_k = -\frac{\xi}{3}(1-n_k^2)\,m_D^2\,;~~\gamma_k=\frac{\xi}{3}\,n_k^2\,m_D^2
\eea

%To obtain the propagator (in the co-variant gauge with Feynman gauge parameter) we invert Eq. (\ref{PIdef}).
For the (Feynman gauge) HL propagator $D_{ij}(q)$ with $\vec q=(0,0,q)$ we obtain, in the static limit,
\bea
D_{ij}(q)=-P_{ij}^{2q}\frac{1}{q^2}-P_{ij}^{1q}\frac{1}{q^2+\alpha_q}.
\eea
For $\xi>0$,
the second term has a space-like pole at $q^2=-\alpha_q>0$, which corresponds to the magnetic Weibel instability \cite{Weibel:1959}. More precisely,
all modes for which $0<q^2<(-\alpha_q)$ exhibit exponential growth with
a growth rate $\gamma(q)$ that approaches zero as $q^2$ approaches 0 or $-\alpha_q$ (see Fig.~\ref{disptran}).
The result for the general static propagator $D_{\mu\nu}(k)$ is given in Ref. \cite{Carrington:2008sp}. To leading order in $\xi$, it has space-like poles at $k^2=-\alpha_k$ and $k^2=-(\alpha_k+\gamma_k)$ determined by (\ref{alphagammaRes}).

\section{Integrand for  $\alpha_q$ at next-to-leading order}
\label{sectionIntegrand}

To obtain ${\rm Im}\,\alpha_{\rm nlo}$ we use (\ref{P1q}) and write:
\bea
\label{togetAl}
{\rm Im}\,\alpha_{\rm nlo}%=P^{\alpha q}_{ij}{\rm Im}\,\Pi_{ij}(q)
=\frac{1}{2}P^{1q}_{ij}(q)\,{\rm Im}\,\Pi_{ij}(q)
\eea
In order to calculate ${\rm Im}\,\Pi^{ij}(\vec q)$ we need the integral corresponding to the first two diagrams in Fig. \ref{staticSE}.
Since the imaginary-time formalism is not available in the nonequilibrium
case we are considering, our starting point is the (resummed) one-loop
expression for the retarded self-energy
as given by the real-time formalism.
% \cite{vanEijck:1994rw}. 
We use the fact that the vertices are pure real in the static limit (see Appendix B) and obtain:
\bea
\label{integrand1}
\Pi^{ij}(\vec q) &=& \frac{i}{4} \,g^2\, C_A \,\int \frac{d^4K}{(2\pi)^4}
\biggl[F^{ij}_{nm}\cdot\left\{D^{sym}_{n}(K) D^{adv}_{m}(R)+D^{adv}_{n}(K) D^{sym}_{m}(R)\right\}\nonumber\\
&&\qquad\qquad +H^{ij}_{n} \cdot\big(D^{sym}_{n}(K)\big)\biggr]
\eea
where we have defined:
\bea
\label{integrandDefns}
&& F^{ij}_{nm}=\big((\Gamma^0+\Gamma)^{\lambda i \tau}P^{nk}_{\lambda\lambda^\prime}(\Gamma^0+\Gamma)^{\lambda^\prime j \tau^\prime} P^{mr}_{\tau\tau^\prime}\big)\bigg|_{q_0=k_0=0} \\[2mm]
&& H^{ij}_{n}=\big((M^0 + M)^{ij\lambda\sigma} P^{nk}_{\lambda\sigma}\big)\bigg|_{q_0=k_0=0} \nonumber\\[2mm]
&&D^{sym}_{n}(K) = \big(1+2n^g_{\rm eq}(k_0)\big)\big(D_n^{ret}(K)-D_n^{adv}(K)\big)\nonumber\\[2mm]
&&D^{sym}_{m}(R) = \big(1+2n^g_{\rm eq}(r_0)\big)\big(D_n^{ret}(R)-D_n^{adv}(R)\big)\nonumber
\eea
The order of the momentum variables is $(K,Q,R)$ for the 3-point functions and $(Q,-Q,K,-K)$ for the 4-point functions.
The repeated indices $n$ and $m$ indicate sums over the projection operators in (\ref{projAll}), and the corresponding mode functions in the propagator. These functions are given in \cite{Carrington:2008sp} and we do not reproduce them here because, as we shall see below, not all components of the propagator contribute at leading order. 

To lowest order in $\xi$ we can use the equilibrium distribution function in the third and fourth lines in (\ref{integrandDefns}).
The dominant contribution to the integral can be extracted by using $(1+2n^g_{\rm eq}(k_0))\to 2T/k_0$ which gives:
\bea
\label{integrand2}
\Pi^{ij}(q_0=0,\vec q) &&= \frac{1}{2} \,g^2\, C_A \,T\, \int \frac{d^3 k}{(2\pi)^3}\; \int\frac{dk_0}{2\pi}\frac{1}{k_0}\\[2mm]
&&\cdot\, \left[\right.F^{ij}_{nm}\cdot \, i\,\big(D^{ret}_{n}(K)D^{ret}_{m}(K+Q)-D^{adv}_{n}(K)D^{adv}_{m}(K+Q)\big) \nonumber\\[2mm]
&& +H^{ij}_{n} \cdot \,i\, \big(D^{ret}_{n}(K)-D^{adv}_{n}(K)\big)\left.\right]\left.\right|_{q_0=0}\nonumber
\eea

Since $D^{ret}=(D^{adv})^*$, this expression appears to be manifestly real.
However, it is not well-defined, because in the anisotropic case, there are
space-like poles in the propagators at $k_0=0$. We discuss this issue in the next section. 

\section{Analytic Structure of the HL Propagator}
\label{sectionProp}

\begin{figure}[H]
\begin{center}
\includegraphics[width=8cm]{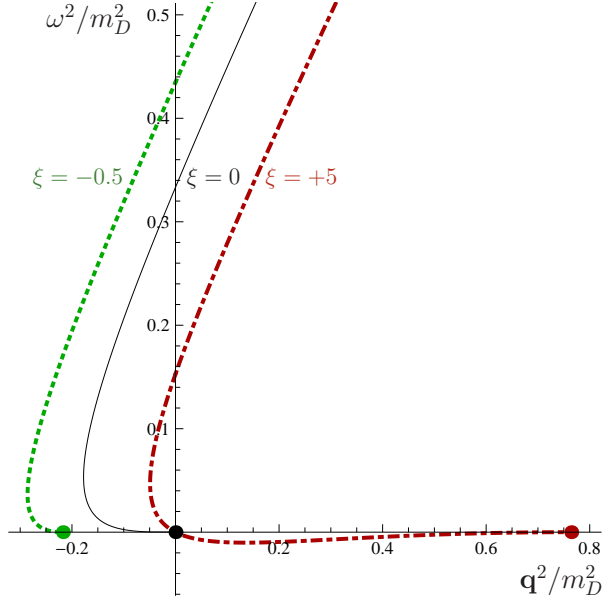}
\end{center}
\caption{Poles in the propagator component $D_\alpha(\omega,\vec q)$ 
for $\vec q$ parallel to the direction of anisotropy, for anisotropy
parameter $\xi=-0.5$ (prolate momentum distribution),
$\xi=0$ (isotropic case), and $\xi=+5$ (oblate case, with
space-like pole at zero frequency).}
\label{disptran}
\end{figure}

The analytic structure of the anisotropic HL propagator defined by
(\ref{PIdef}) has been discussed
in Refs.~\cite{Romatschke:2003ms,Romatschke:2004jh}. 
The poles in the propagator at real frequencies and real wave vectors
determine the dispersion laws of propagating modes. Poles at
real frequencies and imaginary wave vector describe (frequency-dependent)
screening effects. In the anisotropic case, there appear also poles
at real wave vector and imaginary frequency corresponding to plasma
instabilities. As an example,
Fig.~\ref{disptran} displays the location of all these poles in the propagator $D_\alpha(\omega,0,0,q)$ in a plot of
$\omega^2$ versus $k^2$ for 3 values of $\xi$.

%%%%%%%%%%%%%%%%%%%
%%%%%%%%%%%%%%%%%%%%

%The generic location of poles and cuts in the complex frequency plane is summarized in Fig.~\ref{Figdanal}. 

%\par\begin{figure}[H]
%\begin{center}
%\includegraphics[width=6cm]{danal.eps}
%\end{center}
%\caption{Poles and cuts in the anisotropic HL propagator. The poles
%on the imaginary axis appear at sufficiently small $k$ and disappear from 
%the physical sheet at higher $k$.}
%\label{Figdanal}
%\end{figure}

\begin{figure}[H]
\begin{center}
\includegraphics[width=6cm]{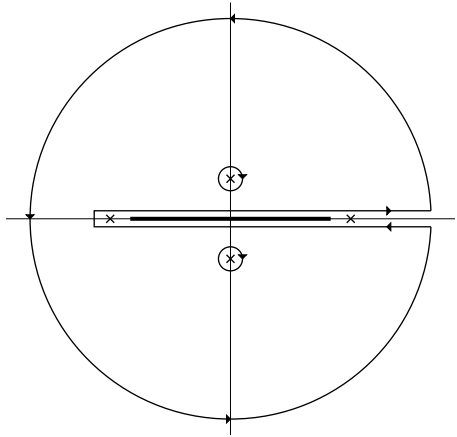}
\end{center}
\caption{Contour used to derive Eq. (\ref{spectral}) and Eq. (\ref{cauchy}). Crosses and thick lines indicate poles and cuts, respectively, in the anisotropic HL propagator. The poles
on the imaginary axis appear at sufficiently small $k$ and disappear from 
the physical sheet at higher $k$. In the case of Eq. (\ref{cauchy}), there may be an additional pair of poles on the imaginary axis from the second propagator in the one-loop diagram that needs to be encircled clockwise.}
\label{figcont1}
\end{figure}

The analytic HL propagator $D_{\mu\nu}(k_0,\vec k)$ with complex $k_0$ but
real $\vec k$
generally has time-like poles on the real axis at $k_0=\omega(k)>k$,
a logarithmic branch cut between $k_0=\pm k$, and, for sufficiently small $k$ (with details depending on the
direction of $\vec k$ and the polarization)
poles at $k_0=\pm i|\gamma(\vec k)|$, where $|\gamma(\vec k)|$ is the growth rate
of unstable modes (not to be confused with the structure function $\gamma_k$ in the decomposition of $\Pi_{ij}(K)$).
When these additional poles are present, using Cauchy's theorem and an integration contour in complex frequency as depicted in Fig.~\ref{figcont1},
the spectral representation of the analytic propagator has the form:
\be
\label{spectral}
D(\omega,\vec k)=\int_{-\infty}^\infty \frac{dk_0}{2\pi i}
\frac{D(k_0+i\epsilon,\vec k)-D(k_0-i\epsilon,\vec k)}{k_0-\omega}
-\sum_{\pm} \frac{{\rm Res}D(\pm i|\gamma(\vec k)|,\vec k)}{\pm i|\gamma(\vec k|)
-\omega}
\ee
The last term in (\ref{spectral}) is produced by the poles at $\pm i\gamma(\vec k)$. In the isotropic case this term is absent and we recover the usual spectral representation. 
\par

However, whenever the analytic propagator $D(z,\vec k)$ has poles on the imaginary axis corresponding to plasma instabilities, 
$D(\omega+i\epsilon,\vec k)$ does no longer yield the retarded propagator,
since it does not satisfy the condition that $\int d\omega e^{-i\omega t} D(\omega+i\epsilon,\vec k)$ vanishes for $t<0$. This point will be discussed in section \ref{sectionMod} where we present a modification of the formalism in which the retarded propagator has the correct causal structure.
In section \ref{sectionNaive} below, we consider the consequences of simply ignoring the issue,
and calculating the next-to-leading order polarization tensor by defining retarded/advanced propagators
as usual through (\ref{Dretadv}).

\section{Next-to-leading order calculation of ${\rm Im}\,\alpha(q_=0,q)$ using unmodified HL resummation}
\label{sectionNaive}

We begin by following the strategy that would work in standard HTL perturbation theory:
We can rewrite the $k_0$-integral in (\ref{integrand2}) using a sum rule technique (see, for example, \cite{LeB:TFT}). For a $k_0'$ away from the real axis, we can
use Cauchy's theorem to write:
\bea
\label{cauchy}
X(k_0^\prime,k)&&=\oint_C\frac{dz}{2\pi i}\,\frac{X(z,k)}{z-k_0^\prime} \\[2mm]
&&=\int^\infty_{-\infty}\frac{dk_0}{2\pi i}\frac{X(k_0+i\epsilon,k)-X(k_0-i\epsilon,k)}{k_0-k_0^\prime}-\sum_{s}\frac{{\rm Res} X(i\gamma_s(k),k)}{i\gamma_s(k)-k_0^\prime}
\nonumber\\
&&=\int^\infty_{-\infty}\frac{dk_0}{2\pi i}\frac{X^{ret}(k_0,k)-X^{adv}(k_0,k)}{k_0-k_0^\prime}-\sum_{s}\frac{{\rm Res}X(i\gamma_s(k),k)}{i\gamma_s(k)-k_0^\prime}\nonumber
\eea
where the contour $C$ is analogous to the one in Fig. \ref{figcont1},
except that there will be generally two pairs of simple poles appearing
symmetrically on the imaginary axis, from 
$D(k_0,\vec k)$ and $D(k_0,\vec r)$, which are denoted collectively by $\gamma_s$.
In the last line we have tentatively assumed the validity of Eq.~(\ref{Dretadv}), which will be reassessed in the next section.
%\par\begin{figure}[H]
%\begin{center}
%\includegraphics[width=6cm]{cont2.eps}
%\end{center}
%\caption{Contour used in Eq. (\ref{cauchy}).}
%\label{contFig}
%\end{figure}

We have to consider $X(k_0,k) = i^2\,H_{n}^{ij}D_n(K)$ and $X(k_0,k) =i^2\, F_{nm}^{ij}D_n(K)D_m(K+Q)$. Taking the limit $k_0^\prime \to 0^+
\equiv \lim_{\epsilon\to0}\lim_{\eta\to0}\eta+i\epsilon$, Eq. (\ref{cauchy}) gives:
\bea
\label{GFrule}
&&{\rm Im\,}\int^\infty_{-\infty}\frac{dk_0}{2\pi}\,\frac{1}{k_0}\, H_{n}^{ij} \,i\,\big(D_n^{ret}(K)-D_n^{adv}(K)\big)=-H_{n}^{ij}\,{\rm Im\,}\lim_{k0\to 0^+} D_n^{ret}(K) \\
&&{\rm Im\,}\int^\infty_{-\infty}\frac{dk_0}{2\pi}\,\frac{1}{k_0}\, F_{nm}^{ij} \,i\,\big(D_n^{ret}(K)D_m^{ret}(K+Q)-D_n^{adv}(K)D_m^{adv}(K+Q)\big)\nonumber\\
&&~~=- F_{nm}^{ij}\,{\rm Im\,}\lim_{k0\to 0^+} D_n^{ret}(K)D_m^{ret}(K+Q)\nonumber
\eea
The second term on the right hand side in (\ref{cauchy}) does not contribute to the imaginary part since the residues
${\rm Res} X(i\gamma(k),k)$ are purely imaginary and odd in $\gamma$.
This means that 
$\sum{{\rm Res} X(i\gamma_s(k),k)}/({i\gamma_s(k)-k_0^\prime})$
is purely real when $k_0^\prime$ approaches the real axis, since
for each value of $\gamma$ there is another with reversed sign. 

Substituting (\ref{GFrule}) into (\ref{integrand2}) we obtain:
\bea
\label{integrand3}
{\rm Im\,}\Pi^{ij}(\vec q) &&= -\frac{1}{2} \,g^2\, C_A %\,\delta_{ab}
\,T\, \\
&&\int \frac{d^3 k}{(2\pi)^3}
\left[\right. F_{nm}^{ij}\,{\rm Im\,}\lim_{k0\to 0^+} D_n^{ret}(K)D_m^{ret}(K+Q)
+H_{n}^{ij}\,{\rm Im\,}\lim_{k0\to 0^+} D_n^{ret}(K)\left.\right]\bigg|_{q_0=0}\nonumber
\eea
%This expression is the starting point
%of the calculation of $\alpha_{nlo}(q_0,\vec q)$ that was proposed in Ref.~\cite{Romatschke:2006bb}. 
The integral in (\ref{integrand3}) could have an
imaginary part coming from the space-like poles of the static internal propagators at $k_0=$ when $\vec k$ is just at the boundary of the instability domain. This observation led to the conjecture in Ref.~\cite{Romatschke:2006bb}
that this imaginary part is in fact nonzero, and that it
regulates the singularities in the momentum broadening coefficient that are produced by the space-like poles in the static HL propagator. 
This conjecture was supported by a partial calculation of the integral in (\ref{integrand3}), in which only the tadpole diagram with a bare vertex was included. Below we calculate the complete result including all HL vertices which are nontrivial in the anisotropic case, in contrast to their HTL counterparts.

To leading order in $\xi$, the poles in the $k$-propagator occur at $k^2=-\alpha_k$ and $k^2=-(\alpha_k+\gamma_k)$, or $k\sim \sqrt{\xi}\,m_D + {\cal O}(\xi)$. Therefore, we need only the part of this propagator that is leading order in $\xi$ for momenta $k\sim \sqrt{\xi}\,m_D$. It is easy to show that only the spatial components contribute.
Since we are working on the LO mass shell $q = \sqrt{\xi/3}\,m_D$, the same conclusion holds for the $r$-propagator. 
Assuming (\ref{Dretadv}), we obtain:
\bea
\label{propLO}
&&\lim_{k_0\to 0^+}D^{ret}_{ij}(K) \\
&&\stackrel{LO}{\to} \lim_{k_0\to 0^+}\bigg[P^{2k}_{ij}\frac{1}{k_0^2-k^2+i\, \mathrm{sgn}(k_0)\epsilon} +P^{3k}_{ij}\frac{1}{k_0^2-(k^2+\alpha_k)+i\, \mathrm{sgn}(k_0)\epsilon}\nonumber\\
&&\qquad\qquad\quad+P^{5k}_{ij}\frac{1}{k_0^2-(k^2+\alpha_k+\gamma_k)+i\, \mathrm{sgn}(k_0)\epsilon}\bigg]\nonumber
\eea
The real and imaginary parts are:
\bea
&&{\rm Re}\left[\right.\lim_{k_0\to 0^+}D^{ret}_{ij}(K)\left.\right] \stackrel{LO}{\to} -P^{2k}_{ij}\frac{1}{k^2} -P^{3k}_{ij}f_3(k)-P^{5k}_{ij}f_5(k) \\[2mm]
&&i\,{\rm Im}\left[\right.\lim_{k_0\to 0^+}D^{ret}_{ij}(K)\left.\right] \stackrel{LO}{\to} -i \pi \big[P^{3k}_{ij}\delta(f_3^{-1})+P^{5k}_{ij}\delta(f_5^{-1})\big] \nonumber\\[2mm]
&& f_3(k) = \frac{1}{k^2+\alpha_k}\,;~~f_5(k) = \frac{1}{k^2+\alpha_k+\gamma_k}\nonumber
\eea
Insertion into (\ref{integrand3}) gives:
\bea
\label{integrand4}
{\rm Im}\,\Pi^{ij}(\vec q) = \frac{1}{2} \,g^2\, C_A %\,\delta_{ab}
\,T\,\pi\, \int \frac{d^3 k}{(2\pi)^3}
\Bigl[&&\sum_
{n,m\in\{3,5\}}
-F_{nm}^{ij}\cdot\big(\delta(f_{nk}^{-1}) f_{mr}+f_{nk} \delta(f_{mr}^{-1})\big)\nonumber\\&&
+\sum_{n\in\{3,5\}}H^{ij}_n\cdot\big(\delta(f_{nk}^{-1})\big)\bigg]
\eea
We obtain ${\rm Im}\alpha_{\rm nlo}$ by substituting (\ref{integrand4}) into (\ref{togetAl}).
We give below the result for the integrand for the tadpole graph. The corresponding expression for the bubble graph is straightforward to obtain, but considerably longer.
\bea
\label{integrandRes}
&&{\rm Im}\,\alpha_{\rm nlo}\bigg|_{tp} = -\frac{1}{4} \,g^2\, C_A \,T\, \pi\,\int \frac{d^3 k}{(2\pi)^3} \\
&& \big[ -M^{iijj}\,\delta(f^{-1}_{3k})-4(k^2+3k_3 q)\frac{1}{k^2}\,\delta(f^{-1}_{3k})-4(k^2-3q(k_3+n_r^2 q/2))\frac{1}{k^2} \delta(f^{-1}_{5k})\big]\nonumber
\eea

In order to perform the integrations, we choose the following representation:
\bea
\label{vectors}
&& \vec n = (0,0,1) \\
&& \vec q= (0,0,q) \nonumber \\
&& \vec k=(k \sin\theta,0,k\cos\theta)\nonumber\\
&& \hat p_1=(\sin \theta_1\cos \phi_1,\sin\theta_1\sin \phi_1,\cos \theta_1)\nonumber\\
&& \hat p_2=(\sin \theta_2\cos \phi_2,\sin\theta_2\sin \phi_2,\cos \theta_2)\nonumber
\eea
The vector $\vec n$ gives the direction of the anisotropy. 
The vectors $\hat p_1$ and $\hat p_2$ are used to calculate the HL vertex components $M^{iijj}$ and $\Gamma^{ijl}\Gamma^{ijl}$. We will write $\cos\theta=x$, $\cos\theta_1=x_1$, $\cos\theta_2=x_2$. We work on the LO mass shell $q = \sqrt{\xi/3}\,m_D$. Using the fact that each term in (\ref{integrandRes}) contains a delta function which gives $k\sim \sqrt{\xi}m_D$, we scale $k$ to obtain a dimensionless variable $\tilde k$ which is defined by: $k= (\sqrt{\xi/3}\,m_D) \, \tilde k$. 
Using this coordinate system, all angular integrals in the HL vertices can be calculated analytically. Next, we use the delta functions to do the polar integral over the variable $x=\cos\theta$. This produces constraints on the $\tilde k$ integration. Finally, the resulting $\tilde k$-integral can be done by numerically. 
The full integrand produces a huge number of terms, many of which are divergent. When all terms are combined, all divergences cancel, which provides a check of our algebra. Some details of the calculation are given in Appendix \ref{AppendixB}. The final result is:
\bea
\label{finalRes}
{\rm Im}(\alpha_q)_{\rm nlo}=-\frac{1}{16 \pi}  \,g^2 \,N_c \,m_D \,T \,\sqrt{\frac{\xi}{3}}\,\cdot\,3.77\ldots
\eea
This finite result confirms 
the conjecture of Ref.~\cite{Romatschke:2006bb} that
a complete one-loop evaluation of
Eq.~(\ref{integrand3}) using full HL vertices and HL propagators
would lead to a nonvanishing imaginary part to the static gluon
polarization tensor that may be used as a regulator for the
space-like poles in the static gluon propagator of an anisotropic plasma.

However, as 
already mentioned in the previous section, the finite result in (\ref{finalRes}) is produced by using HL retarded and advanced propagators which do not have the correct causal structure.
In fact, it is easy to see that there is a fundamental problem with the above calculation. The sign of the imaginary part in (\ref{finalRes}) is determined by having taken the static limit $k_0=\eta+i\epsilon\rightarrow 0^+$ in (\ref{integrand3}). Taking the static limit e.g.\ instead by $k_0=-\eta+i\epsilon$ would produce the opposite sign. 
In the next section we discuss how to define retarded and advanced propagators that have the correct causal structure. We then show that using these propagators produces a pure real result.

\section{Retarded HL propagator and corrected HL resummation}
\label{sectionMod}

\begin{figure}
\begin{center}
\includegraphics[width=6cm]{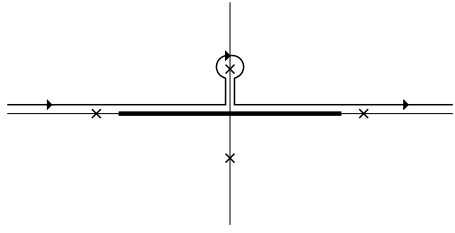}
\end{center}
\caption{Complex frequency contour required to obtain the retarded propagator from the analytic propagator.}
\label{figCr}
\end{figure}

Using the analytic propagator $D$ obtained from Dyson summation
of the HL polarization tensor
according to Eq.~(\ref{PIdef}), we can construct a retarded propagator
\bea
D^{ret}(t,\vec k)=\int_{C_r} d\omega e^{-i\omega t} D(\omega,\vec k)
\eea
with the property that $D^{ret}(t<0,\vec k)\equiv0$ by choosing
a complex contour $C_r$ that 
runs along the real axis and circumvents any poles at positive imaginary
$\omega$ by going over them, as shown in Fig.~\ref{figCr}. 
Moreover, for $t>0$ this propagator 
correctly describes the exponential growth
of the unstable modes, which an integration over just the real axis would
have missed. Restricting
$\omega$ to real values again, we can take the effect of the
contour $C_r$ into account by defining
\be
\label{Dretcorr}
D^{ret}_n(\omega,\vec k)=
D_n(\omega+i\epsilon,\vec k)-\frac{{\rm Res}D_n(+i\gamma_n(\vec k),\vec k)}{\omega+i\epsilon-
i\gamma_n(\vec k)}\Theta(\gamma_n(\vec k)-\epsilon)
\ee
where $D_n$ represents a component of the analytic propagator. 
For $k$ large enough so that all modes are stable, all $\gamma_n(\vec k)$ are
negative and this definition of the retarded propagator coincides with
the usual one. But when $k$ becomes small enough for unstable modes to appear,
the corresponding poles are effectively subtracted.

For example, in the small-$\xi$ limit we have 
\bea
\alpha(0,\vec k)&=&-(\xi/3)m_D^2\cos^2\theta + O(\xi^2) \;
\hbox{with}\; \cos\theta=k^3/k \nonumber\\
\gamma(\vec k)&=&-\frac{4k}{\pi m_D^2}(k^2-(\xi/3)m_D^2\cos^2\theta+O(\xi^2))\nonumber\\
\mathrm{Res}\,D_\alpha(i\gamma,\vec k)&=&\frac{4ik}{\pi m_D^2}+O(\xi)
\eea 
so that the static limit of
the retarded static $\alpha$-propagator responsible
for the Weibel instabilities now reads
\bea
D_\alpha(0,\vec k)=&{\displaystyle 
\frac1{k^2-(\xi/3)m_D^2\cos^2\theta +O(\xi^2)}}&\;
\mbox{for}\;k>\sqrt{\xi/3}m_D\cos\theta\nonumber\\
D_\alpha(0,\vec k)=&O(\xi^0)&\;
\mbox{for}\;k\le\sqrt{\xi/3}m_D\cos\theta.
\eea
In particular, the imaginary part proportional to 
$\delta(k^2+\alpha(0,\vec k))$ is cancelled (irrespective of the
sign of the infinitesimal $\eta$ in $\omega=\eta+i\epsilon$),
and the same is true for all space-like poles in $D$
(and also does not require the simplifying
assumption of momentum parallel to the anisotropy direction used in
our explicit calculations).
Thus the entire
contribution to the imaginary part
calculated in section \ref{sectionNaive} disappears.

%the space-like poles in the static limit of the analytic propagator 
%For $k$ small enough that  $\gamma(\vec k)$ is positive, and unstable space-like poles exist in the static propagator, the second term on the righthand side exactly removes the contribution calculated in section \ref{sectionNaive}. 
%When $k$ is large enough that $\gamma(\vec k)$ is negative, all modes
%are stable, and the second term on the right hand side of (\ref{Dretcorr}) drops out, which reproduces the usual definition of the retarded propagator. 
%In this case however, the space-like pole in the static propagator
%is absent, and the static propagator becomes purely real, which means there is no contribution to the result in \ref{sectionNaive} from this regime.  
%The result is that the imaginary part is identically zero. 

%New conjecture: at least in the limit of small $\xi$ it should
%now be meaningful to use standard real-time perturbation theory
%with HL resummations, if the above retarded propagator is employed.

\section{Conclusion}
\label{sectionConclusions}

%Can we say already something about energy loss and momentum broadening?

%In this paper we have calculated the imaginary part of the next-to-leading static gluon self energy for an anisotropic plasma in the limit of a small momentum space anisotropy.
%We have considered the case where the external momentum is parallel to the anisotropy direction.

%Using the Ward identities for the static hard-loop gluon polarization tensor and the static hard-loop vertices, we have derived a comparatively compact form for the integrand for the structure function containing the space-like pole associated with magnetic instabilities (Eq. (\ref{integrandRes})). We have evaluated the integrals and obtained the finite result given in Eq. (\ref{finalRes}). This calculation is the first complete analytic result at next-to-leading for anisotropic systems.
%The result is important because it regulates the non-integrable singularities that appear in perturbative calculations of jet quenching and momentum broadening. \cite{Romatschke:2006bb,Baier:2008js}.

Starting from the one-loop expressions of the real-time formalism we have
found that these require retarded and advanced propagators, which
in the anisotropic case after resummation of the HL gluon self-energy
are no longer given as boundary values of
the analytic propagator in the limit of real frequencies.
Using the modified prescription (\ref{Dretcorr}) we found that
the imaginary part of the static gluon self-energy vanishes also
at HL-resummed one-loop order. This disproves the conjecture of Ref.~ \cite{Romatschke:2006bb} that such an imaginary part would be generated, and then could be used to regulate the
non-integrable singularities encountered 
 in perturbative calculations of jet quenching and momentum broadening
in an anisotropic plasma \cite{Romatschke:2006bb,Baier:2008js}.

In Ref.~\cite{Baier:2008js} it has been pointed out that the space-like
poles present a problem only together with the singularity of the Bose-Einstein distribution function, and that, as an alternative regularization, it would be natural to assume a lower cutoff
on the frequencies because of the inherently nonequilibrium nature of the
problem, involving characteristic time scales $\sim (g\xi T)^{-1}$.
This resolution to the problem of non-integrable singularities in
the calculations of Ref.~\cite{Romatschke:2006bb,Baier:2008js}
in fact would lead to a larger enhancement of the effects of anisotropy
on jet quenching and momentum broadening than those resulting
from a nonzero imaginary part of the gluon self-energy of
order $g^2 T m_D \sqrt{\xi}$. Our results evidently lend support 
to this possibility.

\acknowledgments

We would like to thank P.\ Romatschke for stimulating discussions.
M.E.C.\ acknowledges financial support from Technische Universit\"at Wien;
A.R.\ acknowledges support from the Austrian Science Foundation, FWF,
project no.\ 19526. 

\appendix
\section{Projection Operators}\label{AppendixA}

The orthogonality relations satisfied by the operators given in section \ref{sectionSE} are:
\bea
&& P^1_{il}P^1_{lj}=P^1_{ij}\,;~~P^1_{il}P^2_{lj}=0\,;~~P^1_{il}P^3_{lj}=P^3_{ij}\,;~~P^1_{il}P^4_{lj}+P^4_{il}P^1_{lj}=P^4_{ij} \\
&&P^2_{il}P^2_{lj}=P^2_{ij}\,;~~P^2_{il}P^3_{lj}=0\,;~~P^2_{il}P^4_{lj}+P^4_{il}P^2_{lj}=P^4_{ij} \nonumber\\
&&P^3_{il}P^3_{lj}=P^3_{ij}\,;~~P^3_{il}P^4_{lj}=0 \nonumber\\
&&P^2_{il}P^5_{lj}=0\,;~~P^3_{il}P^5_{lj}=0\,;~~P^4_{il}P^5_{lj}=P^4_{ij}\,;~~P^5_{il}P^5_{lj} = P^5_{ij}\nonumber\\
&&k_iP^1_{ij}=k_iP^3_{ij}=k_ik_jP^4_{ij}=k_iP^5_{ij}=0 \nonumber\\[4mm]
&&n^k_iP^1_{ij}=n^k_j\,;~~n^k_iP^2_{ij}=n^k_iP^3_{ij}=0\,;~~n^k_i n^k_j P^4_{ij}=0 \,;~~n^k_i P^5_{ij}=n^k_j\nonumber\\
&& {\rm Tr}\,P^1 = 2\,;~~ {\rm Tr}\,P^2 ={\rm Tr}\,P^3 ={\rm Tr}\,P^5= 1\,;~~{\rm Tr}\,P^4=0\nonumber
\eea
\bea
TrP^{1q}=2\,;~~TrP^{2q}=1\,;~~P^{1q}\cdot P^{2q}=0
\eea

\section{Some details of the calculation in section \ref{sectionNaive}}\label{AppendixB}

\subsection{HL vertex components}

The only vertex components that we need are $M^{iijj}$ and $\Gamma^{ijl}\Gamma^{ijl}$.   
Using (\ref{GammaDef}), (\ref{MDef}) and (\ref{vectors}), these vertex components can be rewritten as:

\bea
\label{fints1}
&&M^{iijj}=\frac{1}{4}\int_0^{2\pi} d\phi_1 \int^1_{-1} dx_1 ~\cdot~ I_M, \\
&&I_M= -\frac{6 x}{x x_1^2+\sqrt{1-x^2} \cos \phi_1 \sqrt{1-x_1^2} x_1}+\frac{3 \tilde k x}{\tilde k x x_1^2-x_1^2+\tilde k \sqrt{1-x^2} \cos
   \phi_1 \sqrt{1-x_1^2} x_1}\nonumber\\
   &&~~~~~~~+\frac{3 \tilde k x}{\tilde k x x_1^2+x_1^2+\tilde k \sqrt{1-x^2} \cos \phi_1 \sqrt{1-x_1^2}
   x_1}-\frac{3}{\tilde k x x_1^2-x_1^2+\tilde k \sqrt{1-x^2} \cos \phi_1 \sqrt{1-x_1^2} x_1}\nonumber\\
   &&~~~~~~~+\frac{3}{\tilde k x x_1^2+x_1^2+\tilde k \sqrt{1-x^2} \cos
   \phi_1 \sqrt{1-x_1^2} x_1}, \nonumber\\[2mm]
\label{fints2}
&& \Gamma^{ijl}\Gamma^{ijl}= \frac{1}{4}\int_0^{2\pi} d\phi_1 \int_0^{2\pi} d\phi_2 \int^1_{-1} dx_1 \int^1_{-1} dx_2~\cdot~ I_{\Gamma\Gamma},\\
&& I_{\Gamma\Gamma}=
 \frac{3 x^2 \left(\sqrt{1-x_1^2} \sqrt{1-x_2^2} \cos \left(\phi_1-\phi_2\right)+x_1 x_2\right){}^3}{ \left(\sqrt{1-x^2} \sqrt{1-x_1^2} \cos
   \phi_1+x x_1\right) \left(\sqrt{1-x^2} \sqrt{1-x_2^2} \cos \phi_2+x x_2\right)}\nonumber\\
   &&~~~~~~~-\frac{3 x (\tilde k x+1)
   \left(\sqrt{1-x_1^2} \sqrt{1-x_2^2} \cos \left(\phi_1-\phi_2\right)+x_1 x_2\right){}^3}{ \left(\tilde k \sqrt{1-x^2} \sqrt{1-x_1^2} \cos
   \phi_1+\tilde k x x_1+x_1\right) \left(\sqrt{1-x^2} \sqrt{1-x_2^2} \cos \phi_2+x x_2\right)}\nonumber\\
&&~~~~~~~-\frac{3 x (\tilde k x+1)
   \left(\sqrt{1-x_1^2} \sqrt{1-x_2^2} \cos \left(\phi_1-\phi_2\right)+x_1 x_2\right){}^3}{ \left(\sqrt{1-x^2} \sqrt{1-x_1^2} \cos
   \phi_1+x x_1\right) \left(\tilde k \sqrt{1-x^2} \sqrt{1-x_2^2} \cos \phi_2+\tilde k x x_2+x_2\right)}\nonumber \\
&&~~~~~~~+\frac{3 (\tilde k x+1)^2
   \left(\sqrt{1-x_1^2} \sqrt{1-x_2^2} \cos \left(\phi_1-\phi_2\right)+x_1 x_2\right){}^3}{ \left(\tilde k \sqrt{1-x^2} \sqrt{1-x_1^2} \cos
   \phi_1+\tilde k x x_1+x_1\right) \left(\tilde k \sqrt{1-x^2} \sqrt{1-x_2^2} \cos \phi_2+\tilde k x x_2+x_2\right)}.\nonumber
   \eea
Using the identities $\cos(\phi_1-\phi_2) = \cos \phi_1\cos \phi_2+\sin \phi_1 \sin \phi_2$ and $\sin^2 \phi_1=1-\cos^2 \phi_1$ the azimuthal integrals in Eqs. (\ref{fints1}) and (\ref{fints2}) can be rewritten in the form (with the same expressions for the $\phi_2$ integrals):
 \bea 
 \label{phiInt}
 &&\int \frac{d\phi_1}{2\pi} \frac{1}{A+B \cos \phi_1} = \frac{\text{sgn}(A) \Theta\left(A^2-B^2\right)}{\sqrt{A^2-B^2}}-\frac{i \Theta\left(B^2-A^2\right)}{\sqrt{B^2-A^2}} \\
&&\int \frac{d\phi_1}{2\pi} \frac{\cos \phi_1}{A+B \cos \phi_1} = -\frac{A \text{sgn}(A) \Theta\left(A^2-B^2\right)}{B \sqrt{A^2-B^2}}+\frac{i A \Theta\left(B^2-A^2\right)}{B
   \sqrt{B^2-A^2}}+\frac{1}{B} \nonumber \\
&& \int \frac{d\phi_1}{2\pi}   \frac{\cos ^2\phi_1}{A+B \cos \phi_1} = \frac{\text{sgn}(A) \Theta\left(A^2-B^2\right) A^2}{B^2 \sqrt{A^2-B^2}}-\frac{i \Theta\left(B^2-A^2\right) A^2}{B^2
   \sqrt{B^2-A^2}}-\frac{A}{B^2}\nonumber \\
&& \int \frac{d\phi_1}{2\pi}   \frac{\cos ^3 \phi_1}{A+B \cos \phi_1}=
   -\frac{\text{sgn}(A) \Theta\left(A^2-B^2\right) A^3}{B^3 \sqrt{A^2-B^2}}+\frac{i \Theta\left(B^2-A^2\right) A^3}{B^3
   \sqrt{B^2-A^2}}+\frac{2 A^2+B^2}{2 B^3}\nonumber
   \eea
 
After performing the azimuthal integrals, we are left with the polar integrals over the variables $x_1$ and $x_2$
with limits determined from the theta functions in Eqs.\ (\ref{phiInt}).
%. The limits are determined from the theta functions in Eqs.\ (\ref{phiInt}). 
%The expressions that we need are (with the same expressions for the $x_2$ integrals):
%\bea
%\label{th-ints}
%&& \int dx_1\,\Theta \left(x^2+x_1^2-1\right)\,\cdot\,{\rm fcn}(x_1)=  \int_{-1}^{-x_s(1)}dx_1\,{\rm fcn}(x_1)+\int_{x_s(1)}^{1}dx_1\,{\rm fcn}(x_1) \\
%&& \int dx_1\,\Theta \left(-x^2-x_1^2+1\right)=\int_{-x_s(1)}^{x_s(1)}dx_1\,{\rm fcn}(x_1) \nonumber \\
%&&\int dx_1\,\Theta \left(\left(x^2+x_1^2-1\right) \tilde k^2-2 x x_1^2 \tilde k+x_1^2\right)=\int_{-1}^{-x_s(2)}dx_1\,{\rm fcn}(x_1)+\int_{x_s(2)}^{1}dx_1\,{\rm fcn}(x_1)\nonumber \\
%&& \int dx_1\,\Theta \left(\left(x^2+x_1^2-1\right) \tilde k^2+2 x x_1^2 \tilde k+x_1^2\right)=\int_{-1}^{-x_s(3)}dx_1\,{\rm fcn}(x_1) + \int_{x_s(3)}^{1}dx_1\,{\rm fcn}(x_1)\nonumber \\
%&&\int dx_1\,\Theta \left(-\left(x^2+x_1^2-1\right) \tilde k^2+2 x x_1^2 \tilde k-x_1^2\right)=\int_{-x_s(2)}^{x_s(2)}dx_1\,{\rm fcn}(x_1)\nonumber \\
%&&\int dx_1\, \Theta \left(-\left(x^2+x_1^2-1\right) \tilde k^2-2 x x_1^2 \tilde k-x_1^2\right)=\int_{-x_s(3)}^{x_s(3)}dx_1\,{\rm fcn}(x_1)\nonumber
%\eea
%where 
%\bea
%&& x_s(1)=\sqrt{1-x^2}\,;~~ x_s(2)= \tilde k \sqrt{\frac{1-x^2}{\tilde k^2-2 x \tilde k+1}}\,;~~ x_s(3)= \tilde k
%   \sqrt{\frac{1-x^2}{\tilde k^2+2 x \tilde k+1}}
%   \eea
In every case, the imaginary part contains the integral of an odd function over a symmetric interval and therefore, as is the case for the self energies, the imaginary part of all components is zero in the static limit. This result corresponds to the fact that, in the static limit, the bare propagator has no poles. 
%Using the results in Eq. (\ref{th-ints}), 
The $x_1$- and $x_2$- integrals can be done analytically, yielding
%. The results are:
\be
\label{GammaMres}
  M^{iijj}=0,\quad
  \Gamma^{ijl}\Gamma^{ijl}=
 -\frac{m_D^2 \left(x^2-1\right) \left(\left(8 x^2+7\right) \tilde k^2+\left(8 x^3+22 x\right) \tilde k+8 x^2+7\right) \xi }{3 \left(\tilde k^2+2 x k+1\right)^2}.
\ee

\subsection{Extraction of the imaginary part}

We substitute the results (\ref{GammaMres}) into the integrand and do the $x$-integrals. In order to extract the imaginary part, we write (with corresponding expressions for $f_3(r)$ and $f_5(r)$):
\bea
\label{epsilon}
\pi\delta(f_{3k}^{-1})={\rm Im}\,\frac{1}{k^2+\alpha_k - i \epsilon} \,;~~\pi\delta(f_{5k}^{-1})={\rm Im}\,\frac{1}{k^2+\alpha_k+\gamma_k - i \epsilon}
\eea
Using Eqs.\ (\ref{alphagammaRes}) we can rewrite these expressions:
\bea
\label{expansions}
&& \frac{1}{k^2+\alpha_k - i \epsilon}=-\frac{3}{\xi m_D^2 (x-i\epsilon -x_{3k}^-) (x+i\epsilon -x_{3k}^+)}\\
&& \frac{1}{r^2+\alpha_r - i \epsilon}=\frac{\tilde k^2+2 x \tilde k+1}{\tilde k^2 \xi m_D^2  (x+i\epsilon -x_{3r}^+)(x-i\epsilon -x_{3r}^-)}\nonumber \\
&&\frac{1}{k^2+\alpha_k+\gamma_k - i \epsilon}=-\frac{3}{2 \xi m_D^2 (x-i\epsilon -x_{5k}^-) (x+i\epsilon -x_{5k}^+)}\nonumber \\
&& \frac{1}{r^2+\alpha_r+\gamma_r - i \epsilon}=\frac{3 \left(\tilde k^2+2 x \tilde k+1\right)}{2 \tilde k^2 \xi m_D^2 (x+i\epsilon -x_{5r}^+) (x-i\epsilon -x_{5r}^-)}\nonumber
\eea
with 
\bea
&& x_{3k}^- = -\tilde k\,;~~x_{3k}^+= \tilde k\\
&& x_{3r}^-= -\tilde k\,;~~ x_{3r}^+= -\frac{\tilde k^2+2}{3\tilde k}\quad
\text{when}\; \tilde k<1 \nonumber\\
&& x_{3r}^+= -\tilde k\,;~~ x_{3r}^-= -\frac{\tilde k^2+2}{3\tilde k}\quad
\text{when}\; \tilde k>1 \nonumber\\
&& x_{5k}^-= -\frac{\sqrt{\tilde k^2+1}}{\sqrt{2}}\,;~~ x_{5k}^+= \frac{\sqrt{\tilde k^2+1}}{\sqrt{2}}\nonumber\\
&& x_{5r}^+=
   -\frac{2 \tilde k^2+\sqrt{2 \tilde k^4-2 \tilde k^2+1}+1}{2 \tilde k}\,;~~ x_{5r}^-= \frac{-2 \tilde k^2+\sqrt{2
   \tilde k^4-2 \tilde k^2+1}-1}{2 \tilde k}\nonumber
\eea
Substituting (\ref{epsilon}) and (\ref{expansions}) into (\ref{integrandRes}) 
the $x$-integrals can be done analytically. After a partial fraction expansion, the integration of each term produces either a logarithm, or something that is pure real. The imaginary part of each logarithm gives rise to a theta function, restricting the remaining $\tilde k$-integrals to intervals $0< \tilde k < 1$, $1< \tilde k < 2$, or $0< \tilde k < 2$. 
We note that particular care must be taken because of the double pole at $x_{3k}^-=x_{3r}^-=-\tilde k$
that occurs when $\tilde k < 1$. We regulate this term in the usual way by writing:
\bea
\frac{1}{(x-x_{3k}^-)(x-x_{3k}^-)} = \frac{d}{dM}\frac{1}{x-x_{3k}^--M}
\eea
and taking $M$ to zero at the end of the calculation.

%\subsection{$\tilde k$-integrals}

Once all the $x$-integrals are done, the remaining integral over $\tilde k$ can be done numerically. There are many terms that are divergent at $\tilde k=0$, $\tilde k=1$, and/or $\tilde k=2$. Summing all contributions yields the result
given in Eq.~(\ref{finalRes}).
%\bea
%\label{finalResap}
%{\rm Im}(\alpha_q)_{\rm nlo}=-\frac{1}{16 \pi}  \,g^2 \,N_c \,m_D \,T \,\sqrt{\frac{\xi}{3}}\,\cdot\,3.77\ldots
%\eea

%\bibliographystyle{prsty}
%\bibliography{ar,tft,qft,books}

%\end{document}

\end{document}